\documentclass[aps,prb,preprint,superscriptaddress]{revtex4-2}

\usepackage{dcolumn}   
\usepackage{bm}        
\usepackage{amsmath}
\usepackage{amssymb}
\usepackage[utf8]{inputenc}
\usepackage{graphicx,siunitx}
\usepackage{amsmath}
\DeclareUnicodeCharacter{2212}{-}
\usepackage{booktabs}
\usepackage{subcaption}
\usepackage[svgnames]{xcolor}
\usepackage[export]{adjustbox}
\usepackage[percent]{overpic}
\usepackage{array}
\usepackage[style=base]{caption}
\DeclareSIUnit\angstrom{\text {Å}}
\begin{document}
	\title{Pressure induced ferromagnetic to antiferromagnetic phase transition in transition metal chalcogenide Cr$_{3}$Te$_4$}
	
	\author{Asish Kumar Mishra}
	\affiliation{Department of Physical Sciences, Indian Institute of Science Education and Research Kolkata, Mohanpur Campus, Mohanpur 741246, Nadia, West Bengal, India.}
	\affiliation{National Centre for High-Pressure Studies, Department of Physical Sciences, Indian Institute of Science Education and Research Kolkata, Mohanpur Campus, Mohanpur 741246, Nadia, West Bengal, India.}
	\author{Souvick Chakraborty}
	\affiliation{Department of Physical Sciences, Indian Institute of Science Education and Research Kolkata, Mohanpur Campus, Mohanpur 741246, Nadia, West Bengal, India.}
	\author{Bidisha Mukherjee}
	\affiliation{Department of Physical Sciences, Indian Institute of Science Education and Research Kolkata, Mohanpur Campus, Mohanpur 741246, Nadia, West Bengal, India.}
	\affiliation{National Centre for High-Pressure Studies, Department of Physical Sciences, Indian Institute of Science Education and Research Kolkata, Mohanpur Campus, Mohanpur 741246, Nadia, West Bengal, India.}
	\author{Mrinmay Sahu}
	\affiliation{Department of Physical Sciences, Indian Institute of Science Education and Research Kolkata, Mohanpur Campus, Mohanpur 741246, Nadia, West Bengal, India.}
	\affiliation{National Centre for High-Pressure Studies, Department of Physical Sciences, Indian Institute of Science Education and Research Kolkata, Mohanpur Campus, Mohanpur 741246, Nadia, West Bengal, India.}
	\author{Suvashree Mukherjee}
	\affiliation{Department of Physical Sciences, Indian Institute of Science Education and Research Kolkata, Mohanpur Campus, Mohanpur 741246, Nadia, West Bengal, India.}
	\affiliation{National Centre for High-Pressure Studies, Department of Physical Sciences, Indian Institute of Science Education and Research Kolkata, Mohanpur Campus, Mohanpur 741246, Nadia, West Bengal, India.}
	\author{Shubham Purwar}
	\affiliation{Department of Condensed Matter and Materials Physics, S. N. Bose National Centre for Basic Sciences, Kolkata, West Bengal 700106, India}
	\author{Harekrishna Bhunia}
	\affiliation{Department of Physical Sciences, Indian Institute of Science Education and Research Kolkata, Mohanpur Campus, Mohanpur 741246, Nadia, West Bengal, India.}
	\author{S. Thirupathaiah}
	\affiliation{Department of Condensed Matter and Materials Physics, S. N. Bose National Centre for Basic Sciences, Kolkata, West Bengal 700106, India}
	\author{Peter Liermann}
	\affiliation{Photon Science, Deutsches Elektronen Synchrotron, 22607 Hamburg, Germany}
	\author{Satyabrata Raj}
	\affiliation{Department of Physical Sciences, Indian Institute of Science Education and Research Kolkata, Mohanpur Campus, Mohanpur 741246, Nadia, West Bengal, India.}
	\affiliation{National Centre for High-Pressure Studies, Department of Physical Sciences, Indian Institute of Science Education and Research Kolkata, Mohanpur Campus, Mohanpur 741246, Nadia, West Bengal, India.}
	\author{Goutam Dev Mukherjee}
	\email [Corresponding author: ]{goutamdev@iiserkol.ac.in}
	\affiliation{Department of Physical Sciences, Indian Institute of Science Education and Research Kolkata, Mohanpur Campus, Mohanpur 741246, Nadia, West Bengal, India.}
	\affiliation{National Centre for High-Pressure Studies, Department of Physical Sciences, Indian Institute of Science Education and Research Kolkata, Mohanpur Campus, Mohanpur 741246, Nadia, West Bengal, India.}
	\date{\today}

	\begin{abstract}
	We have carried out a detailed high-pressure investigation on the strongly correlated transition metal chalcogenide $Cr_{3}Te_4$ using Raman spectroscopy and XRD, which is ferromagnetic and metallic at ambient conditions. We find that the monoclinic structure remains stable up to 30 GPa, the highest pressure studied. The Cr-Te bond length and octahedral volume decrease drastically up to 7.6 GPa pressure. The $A_{1g}$ Raman mode shows a red shift up to 7.6 GPa, and the $E_g$ Raman mode shows a sudden drop around the same pressure. Further low-temperature Raman spectroscopic investigation shows that the Raman modes soften at the ferromagnetic to antiferromagnetic phase transition. This suggests a change in the magnetic ordering at high pressure. Our Density Functional Theory (DFT) calculations reveal the change in magnetic ground state from ferromagnetic state to antiferromagnetic state above 7.6 GPa pressure, corroborating our experimental result.      
		
	\end{abstract} 
	\maketitle
	\newpage
	\section{Introduction}
	Strongly correlated transition metal chalcogenides (TMC) have emerged as one of the most interesting compounds for the science community due to their diverse physical properties \cite{huang2021two}. The presence of room temperature ferromagnetism, 2D superconductivity, metal-insulator transition, and enhanced optoelectronic properties makes these materials ideal \cite{xia2021ferromagnetism,cai2022room,coleman1981superconductivity,duvjir2018emergence,deng2020strong}. Among these materials, $Cr_{1-\delta}Te$ based compounds are of considerable interest. These materials have repeating layers of Cr occupied and Cr vacant sites along the c-axis. The structural, electronic, and magnetic properties in these compounds are sensitive to the $\delta$ value  \cite{ipser1983transition,shimada1996photoemission}. The variation in the $\delta$ value for Cr stoichiometry leads to the stabilization of different crystal structures in this family. The parent compound of this family, CrTe ($\delta$=0), crystallizes in the hexagonal structure \cite{shelke2025electron}. As  Cr concentration decreases, several new crystal structures are formed. For example $Cr_{2}Te_{3}$ ($\delta$=0.34) crystallizes in the Trigonal structure, $Cr_{3}Te_{4}$ ($\delta$=0.25) is found in the monoclinic structure and $Cr_{5}Te_{8}$ ($\delta$=0.37) attains trigonal or monoclinic structure depending on synthesis condition \cite{bensch1997determination,PhysRevB.109.054413,purwar20243d}. The Curie temperature also varies with  Cr concentration with $T_C$ in the range of $ 170 -360 K$ \cite{dijkstra1989band,ipser1983transition,shimada1996photoemission}. These kinds of changes in the structural and magnetic properties may also be induced by externally perturbing the system by applying pressure and temperature. Pressure is an important thermodynamic parameter that has been found to tune the structural, electronic, and magnetic properties of TMCs \cite{ghosh2022structural,saha2020pressure,ishizuka2001pressure,srivastava1969pressure}.
	
	Pressure-induced structural, electronic, and magnetic transitions are observed in several strongly correlated materials. Ishizuka et al. reported the disappearance of ferromagnetic (FM) phase in $Cr_{48}Te_{52}$ ($Cr_{0.93}Te$) above 7 GPa, which is not related to any structural change \cite{ishizuka2001pressure}. In the van der Walls(vdW) ferromagnetic material $Cr_{2}Ge_{2}Te_{6}$, a spin state transition is observed with the application of pressure where the easy axis of magnetization changes from the c axis to the ab plane \cite{lin2018pressure}. A modification in the next nearest neighbour exchange interaction is observed with the application of pressure in $Lu_{2}NiMnO_{6}$ that drives the system to an antiferromagnetic (AFM) phase from a ferromagnetic ground state \cite{terada2020pressure}. Among the TMCs, $Cr_{3}Te_{4}$ is a layered compound, which is ferromagnetic at ambient conditions. It undergoes two magnetic phase transitions: being paramagnetic above 322 K and canted antiferromagnetic below 100 K \cite{PhysRevB.109.054413}. Low temperature electrical resistivity measurements show the metallic nature of the sample  \cite{PhysRevMaterials.7.094204}. The multiple magnetic phases in the sample arise from the direct exchange, intra-layer super exchange, and inter-layer double exchange interaction present in the sample \cite{wang2023field}. Pressure can effectively modulate the inter- and intra-layer distances that can change the magnetic exchange interaction pathway. It can also significantly influence the structural and electronic properties.

	In the present work, we have carried out a detailed high-pressure and low-temperature study on $Cr_{3}Te_4$. The high-pressure investigation includes X-ray diffraction (XRD) and Raman spectroscopic measurements. Low-temperature Raman scattering measurements were used to observe the effect of magnetic ordering on the lattice vibration. High-pressure XRD reveals the absence of a structural phase transition in the sample. The anomalies observed in the Cr-Te bond length and octahedral volume indicate towards the possibility of a pressure-induced change in the magnetic ground state. These changes are also reflected in Raman spectroscopy data, where the $A_{1g}$ and $E_g$ Raman modes show anomalies around the same pressure. The low-temperature Raman data confirm that the Raman mode frequency softens at the ferromagnetic to antiferromagnetic transition.  Density Functional Theory (DFT) calculations also predict a ferromagnetic to antiferromagnetic transition above 7.6 GPa pressure. Our high-pressure XRD, Raman scattering, low-temperature Raman scattering measurements, and the DFT calculations predict a pressure-induced ferromagnetic to antiferromagnetic phase transition.    
	\section{Experimental section}
    High-purity single crystals of $Cr_{3}Te_{4}$ were synthesized by using the chemical vapor transport method as per the procedure described in Ref.\cite{PhysRevMaterials.7.094204}. The phase purity of the sample was verified using XRD, Energy Dispersive X-ray (EDX), and Raman spectroscopic measurement. EDX data suggest the chemical composition of the as-grown crystals to be $Cr_{3.04}Te_4$, which is very close to the desired nominal composition of $Cr_3Te_4$. (Fig.S1 of supplementary material) 
    
     High-pressure XRD measurements were carried out at the PETRA III, P02.2 beamline at the wavelength $\sim$ 0.2907 \AA,  and with an XRD beam spot size of (8$\times$3)~\SI{}{\micro\meter}$^2$. A small portion of the single crystal was crushed into fine powder and loaded into the symmetric diamond anvil cell along with a ruby ball. The ruby fluorescence technique was used for pressure calibration \cite{mao1986calibration}. Neon gas was used as a pressure transmitting medium (PTM). The sample-to-detector distance was calibrated by using the XRD pattern of standard material CeO$_2$. DIOPTAS software \cite{Prescher03072015} was used to convert the 2D diffracted image into $2\theta$ versus intensity profile. GSAS software \cite{Toby:hw0089} was used for the Rietveld refinement of the observed XRD data.

    High-pressure Raman spectroscopic measurements were performed using a piston-cylinder type diamond anvil cell (DAC) from Almax EasyLab Co. (UK) having a culet diameter of \SI{300}{\micro\meter}. A stainless steel gasket of \SI{280}{\micro\meter} thickness was preindented to \SI{45}{\micro\meter} thickness. A central hole of diameter \SI{100}{\micro\meter} was drilled using an electric discharge machine (EDM). Few ruby chips are loaded along with the sample for the purpose of pressure calibration by the Ruby fluorescence technique \cite{mao1986calibration}. $4:1$ methanol-ethanol mixture was used as a pressure transmitting medium (PTM). Pressure-dependent Raman spectra were taken using a confocal micro Raman spectrometer(Horiba Jobin-Yvon LabRam HR-800)in backscattering geometry equipped with \SI{488}{\nano\meter} laser. An infinitely corrected long working distance 20X objective was used to focus the laser beam with spot size $\sim$ \SI{5}{\micro\meter}. Extreme care was taken to collect the Raman spectra from the center of the culet. 
    
    The low-temperature Raman spectroscopic measurements were carried out by using a closed-cycle cryostat equipped with a monovista Raman spectrometer from S\&I GmbH. The sample was positioned on a copper sample holder, attached to the cold head. Lakeshore 325 temperature controller was used to control the temperature, and a Cernox sensor placed very close to the sample measured the temperature of the sample. The Raman spectra were collected in the back-scattering geometry using a 532 nm laser.
    \section{COMPUTATIONAL DETAILS}
   Density Functional Theory (DFT) calculations were carried out using Vienna Ab Initio Simulation Package (VASP) \cite{PhysRevB.54.11169} using the generalized gradient approximation exchange-correlation functional in the form of the Perdew-Burke-Ernzerhof (GGA-PBE) \cite{PhysRevLett.77.3865}. The frozen-core Projector Augmented Wave (PAW) \cite{PhysRevB.50.17953} technique with an energy cutoff of 520 eV for the plane wave basis was chosen for the calculations. The lattice parameters and the atomic positions used in the calculation were taken from the experimental data for different values of pressure. For all the calculations, the energy convergence tolerance was set at $10^{-8}$ eV. There are total 14 atoms (6 Cr and 8 Te) in a single unit cell of $Cr_{3}Te_4$. Different magnetic states (AFM and FM) were considered by switching the magnetic moments of the 6 Cr atoms. A gamma-centred k-point grid of $6\times18\times12$ was used during the self-consistent field calculations for all the calculations.
    \section{RESULTS AND DISCUSSION}
    \subsection{Ambient and high pressure XRD}
\begin{figure}[ht!]
		\centering
		\includegraphics[width=1.0\linewidth]{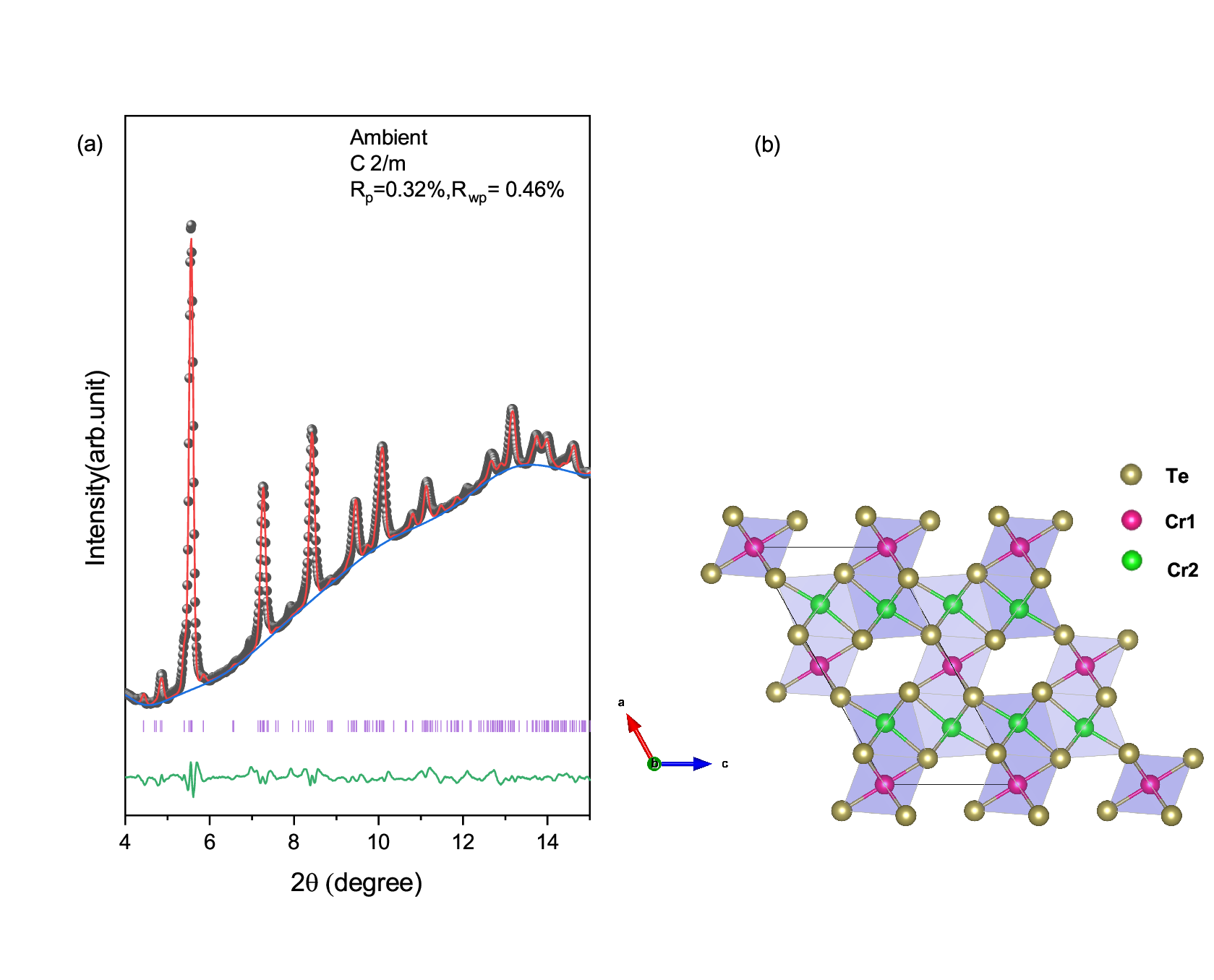}
		\caption{(a) The Rietveld refinement of the ambient XRD pattern (b) Octahedral as well as ball and stick representation of the unit cell of $Cr_{3}Te_4$. The golden circle represents the Te atom, pink and green circle represents $Cr1$ and $Cr2$ respectively}
		\label{Amb XRD}
	\end{figure}

    The ambient XRD pattern of $Cr_{3}Te_{4}$ is indexed to the monoclinic structure in space group $C 2/m$. The obtained lattice parameters are, $a=14.099(2)~\si{\angstrom}$, $b=3.9488(3)~\si{\angstrom}$, $c=6.903(1)~\si{\angstrom}$, $\beta=118.73(1)$, and the unit cell volume is $337.01(8)~\si{\angstrom}^3$. These values match well with those reported in the literature \cite{PhysRevMaterials.7.094204,PhysRevB.109.054413}. We have carried out the Rietveld refinement by taking the initial atomic positions from Goswami et al. \cite{PhysRevB.109.054413}. The Rietveld refinement of the ambient data is shown in Fig.~\ref{Amb XRD}(a), which gives a very good fit with $R_P=0.32\%$ and $R_wp=0.46\%$. The refined atomic positions are given in Table~\ref {table 1}.
    \begin{table}[ht!]
    	\begin{center}
    	\begin{tabular}{|c|c|c|c|c|}
    		\hline
    		Atom&Site&x/a&y/b&z/c\\
    		\hline
    		Cr$(1)$&2a&0.0000&0.0000&0.0000\\
    		\hline
    		Cr$(2)$&4i&0.260(2)&0.0000&0.260(5)\\
    		\hline
    		Te$(1)$&4i&0.3694(8)&0.0000&0.032(2)\\
    		\hline
    		Te$(2)$&4i&0.1105(6)&0.0000&0.431(1)\\
    		\hline
       	\end{tabular}
    	\end{center}
    	\caption{The relative atomic positions obtained from Rietveld refinement at ambient pressure}
    	\label{table 1}
    \end{table}
    The unit cell of $Cr_{3}Te_4$ is shown in Fig.~\ref*{Amb XRD}(b), which is formed by edge sharing and face sharing $CrTe_6$ octahedra. It is a layered structure with alternating Cr filled and Cr vacant layers along the c-axis.

\begin{figure}[ht!]
		\centering
		\includegraphics[width=0.95\linewidth]{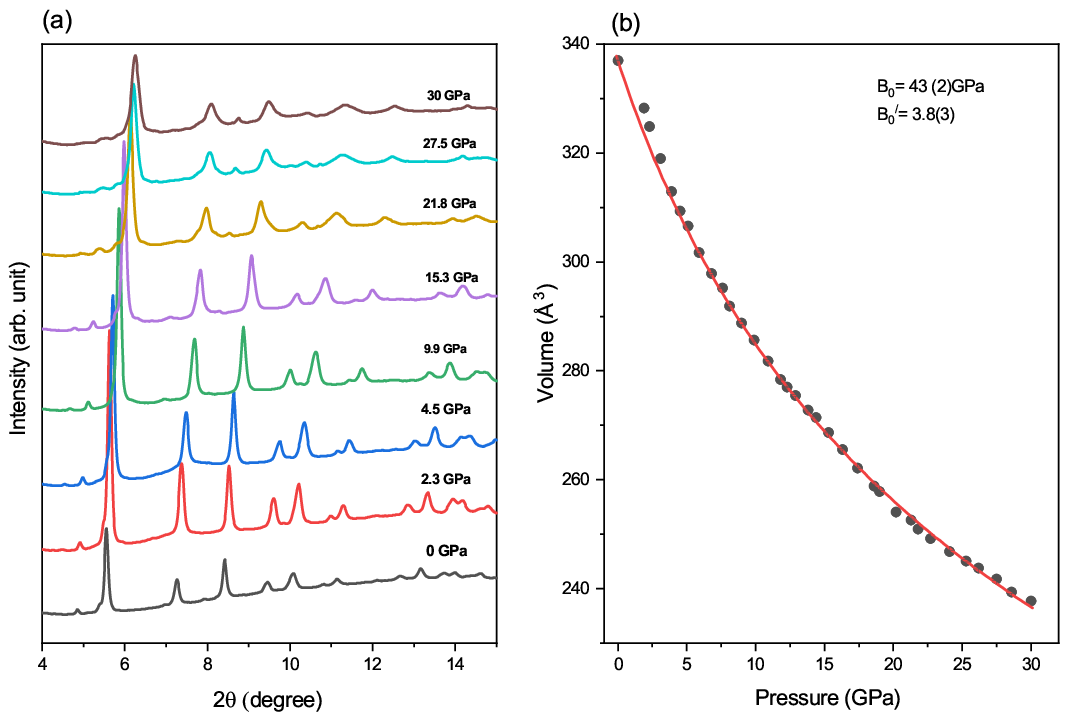}
		\caption{(a) The Pressure evolution of XRD pattern at some selected pressure points (b) Variation of unit cell volume with pressure. The black dots are the experimental data and the red curve is the $3^{rd}$ order BM EOS fit to the experimental data. The symbol sizes denote the errors in volume measurement.  }
		\label{PV variation XRD}
	\end{figure}

    We have collected the XRD patterns of $Cr_{3}Te_4$ at different pressures up to 30 GPa. The pressure evolution of the XRD pattern at certain selected pressure points is shown in Fig.~\ref{PV variation XRD}(a). No change appears with the increase in pressure, and all the XRD patterns can be fitted to the monoclinic structure in space group $C 2/m$ up to the highest pressure studied. This reveals the absence of any structural phase transition up to about 30 GPa. Now, in order to understand the effect of pressure on the structural properties, we have performed a detailed Rietveld analysis. Pressure-dependent variation of the relative lattice parameters reveals that the c-axis becomes more compressible as compared to the a-axis beyond 8 GPa, indicating anisotropic lattice compression as shown in Fig.S2 of the supplementary material. But the pressure-volume curve does not show any significant change around that pressure region, also the structural symmetry remains the same. The unit cell volume obtained after Rietveld refinement is plotted against pressure as shown in Fig.~\ref{PV variation XRD}(b). The 3$^{rd}$ order Birch-Murnaghan (BM) equation of state (EoS)\cite{PhysRev.71.809,doi:10.1073/pnas.30.9.244} is used to fit the experimental data.
     \begin{equation}
    	P(V)= \frac{3B_0}{2}\left[(\frac{V_0}{V})^{7/3}-(\frac{V_0}{V})^{5/3}\right]\times\left\{1+\frac{3}{4}(B^\prime_0-4)\left[(\frac{V_0}{V})^{2/3}-1\right]\right \}
    \end{equation}
     where $B_0$ is the isothermal bulk modulus and $B^\prime_0$ is the first order pressure derivative of $B_0$, $V_0$ is the ambient volume. The experimental data were fitted to the above equation using Eosfit7\cite{Gonzalez-Platas:kc5039}. The value of bulk modulus ($B_0$) is 43(2)~GPa and the first derivative of bulk modulus ($B^\prime_0$) is 3.8(3). The obtained value of bulk modulus is comparable with the other members of the Cr-Te family \cite{eto2001pressure}. Li et al. reported a higher value of bulk modulus of 72.6(2) GPa for $Cr_{3}Te_4$ \cite{li2022pressure}. This discrepancy may originate from the differences in experimental conditions, particularly the choice of pressure-transmitting medium (PTM). In their high-pressure XRD measurements, they have used Silicone oil as a PTM, which remains hydrostatic only up to approximately 3.4 GPa. In our study, we used neon gas, which maintains a hydrostatic condition up to 15 GPa. The increased non-hydrostaticity in their experiment may have led to an overestimation of the bulk modulus. Also in their EOS fit, they have treated the zero pressure volume ($V_0$) as a free parameter, whereas in our fitting, we have fixed the $V_0$ to the experimentally observed value. This difference in fitting procedure could also contribute to the observed variation in the bulk modulus values. 
     
\begin{figure}[ht!]
		\centering
		\includegraphics[width=0.95\linewidth]{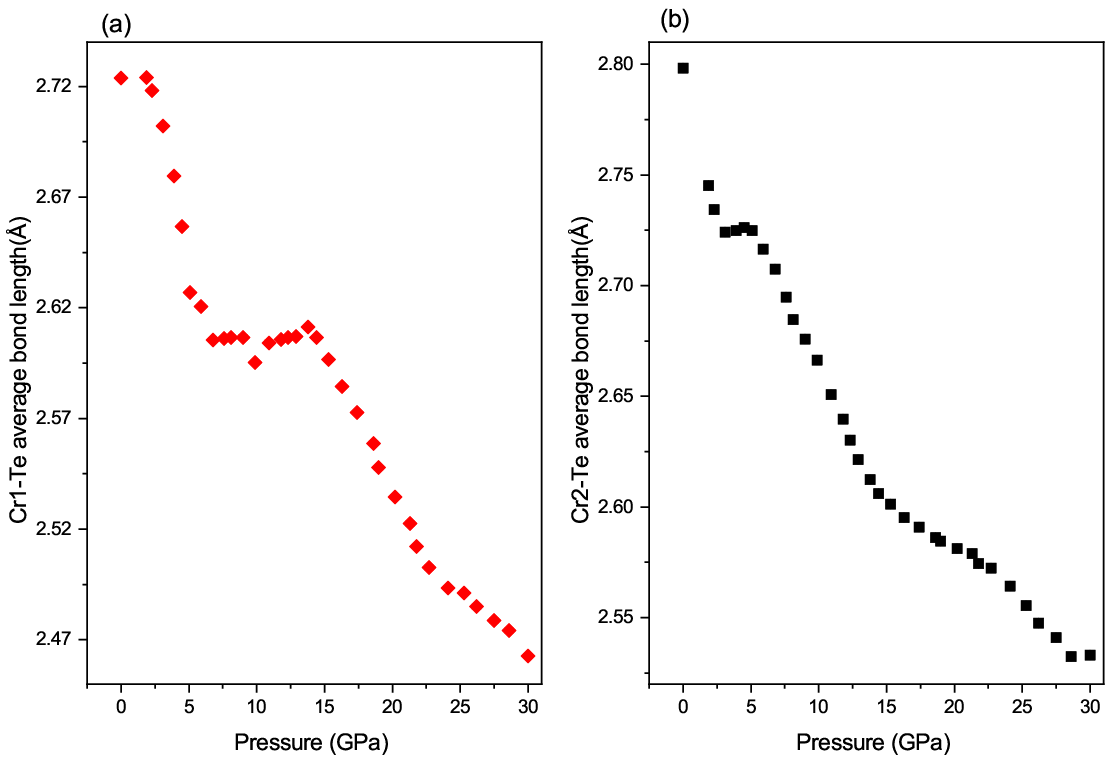}
		\caption{(a) Variation of Cr1-Te bond length with pressure (b) Variation of Cr2-Te bondlength with Pressure}
		\label{bondlength}
		
	\end{figure}

      The unit cell arrangement of  $Cr_{3}Te_4$  offers an interesting perspective. There are two distinct types of octahedra present in the system, and we term them as $Cr_1Te_6$ and $Cr_2Te_6 $. Typically, the volume of octahedra decreases with increasing pressure. $Cr_1Te_6$ octahedral volume decreases by about 27\:\% up to 30 GPa, where as for $Cr_2Te_6$ it is about 29\:\% (Fig. S3 of supplementary material). The values are comparable to the total volume decrease of about 29\:\% . This shows that the octahedral compressibility influences the total volume compression. For a closer inspection, we follow up the pressure evolution of the Cr-Te bond length (See Fig.~\ref{bondlength}). Interestingly, we find that $Cr_1-Te$ bond length reduces drastically up to about 7.6 GPa, and then remains almost constant up to 15 GPa, followed by a regular decrease. $Cr_2-Te$ bond length drops till about 3.5 GPa and again decreases gradually after remaining stable up to about 5.2 GPa. This anomalous behavior of Cr-Te bond lengths with pressure is related to the Cr vacancy in the unit cell. It is already known that the magnetic behavior of the Cr-Te system is influenced by the Cr-Te bond \cite{wang2023field}. Therefore, in the absence of any structural transition, this anisotropic behavior of lattice parameters as well as the anomalous behavior of $CrTe_6$ octahedra are quite interesting and may be due to a pressure-induced modification in the material's electronic or magnetic properties. To study the influence of pressure on the lattice dynamics and to find the correlation of these structural rearrangements with the material's property under pressure, we have carried out high high-pressure Raman spectroscopic study.

    \subsection{High pressure Raman}
\begin{figure}[ht!]
		\centering
		\includegraphics[width=0.85\linewidth]{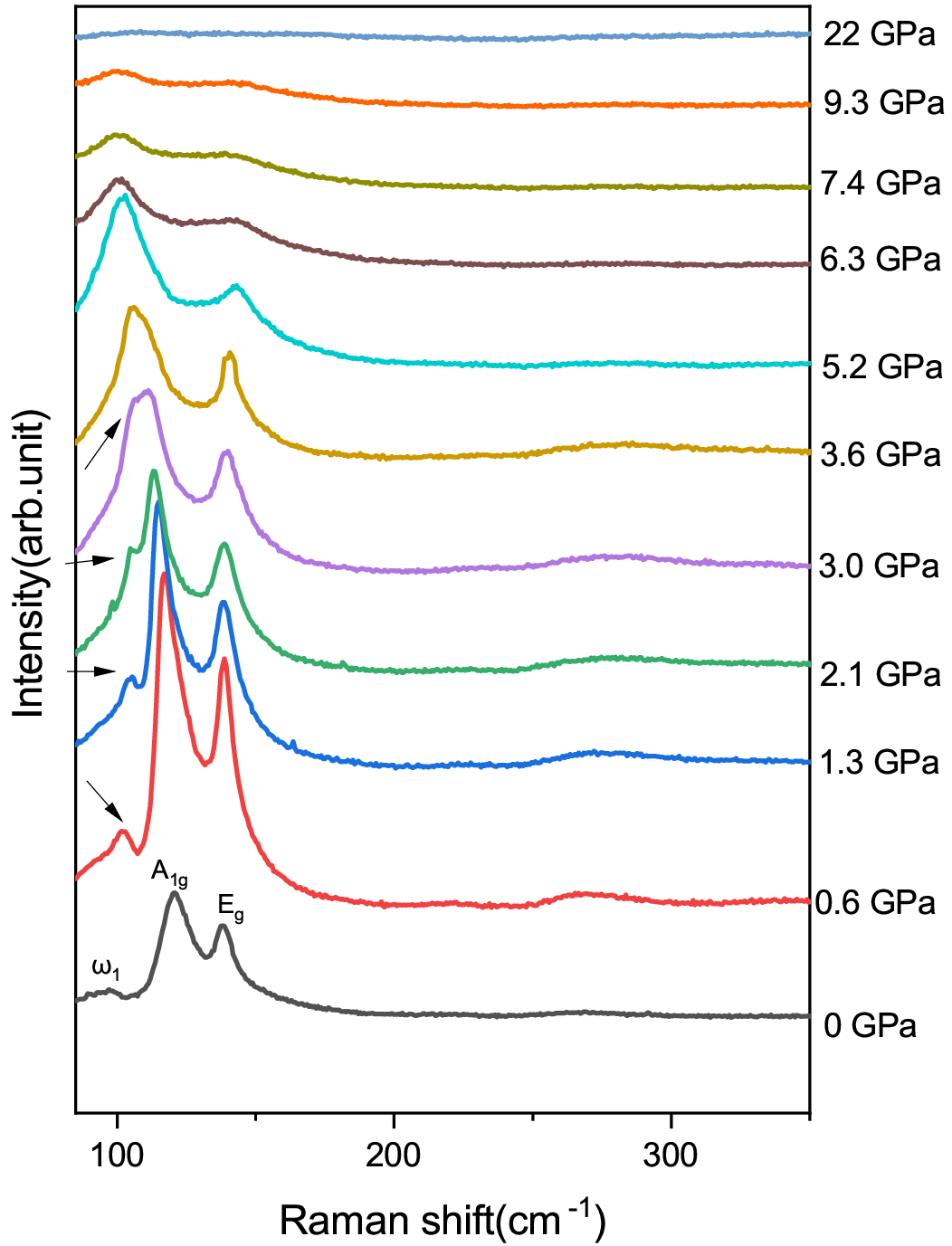}
		\caption{The pressure evolution of Raman spectra at some selected pressure points. The black arrow is a guide to the eye to observe the shift of $\omega_1$ mode with pressure }
		\label{raman pressure evolution}
	\end{figure}

     The pressure evolution of Raman spectra at some selected pressure values is shown in Fig.~\ref{raman pressure evolution}. The ambient Raman spectrum consists of four observable Raman modes as shown in Fig. S4(a) of the supplementary material, which matches well with the previously reported data. The Lorentzian profile is used to fit the background-corrected spectrum. The two prominent Raman modes centered around 120 $cm^{-1}$ and 139 $cm^{-1}$ are assigned to $A_{1g}$ mode and $E_g$ mode respectively, which arise due to the out of plane and the in plane vibrations of the Te atom \cite{D2NA00835A}. The other two modes centered around 95 $cm^{-1}$ and 274 $cm^{-1}$ are denoted as $\omega_1$ and $\omega_2$, respectively. Interestingly, we find $A_{1g}$ mode softens with pressure. The other three modes show a blue shift with an increase in pressure. At about 5.2 GPa pressure, both the  $\omega_1$ and $A_{1g}$ modes merge with each other. With the increase in pressure, the intensity of all the Raman modes decreases. Above 15 GPa the intensity of all the modes becomes very low, so the spectra looks like a flat line. The flatness of Raman spectra can not be assigned to amorphization or disorder in the sample, as XRD Bragg peaks remain well defined at these pressures. 
     
\begin{figure}[ht!]
		\centering
		\includegraphics[width=0.65\linewidth]{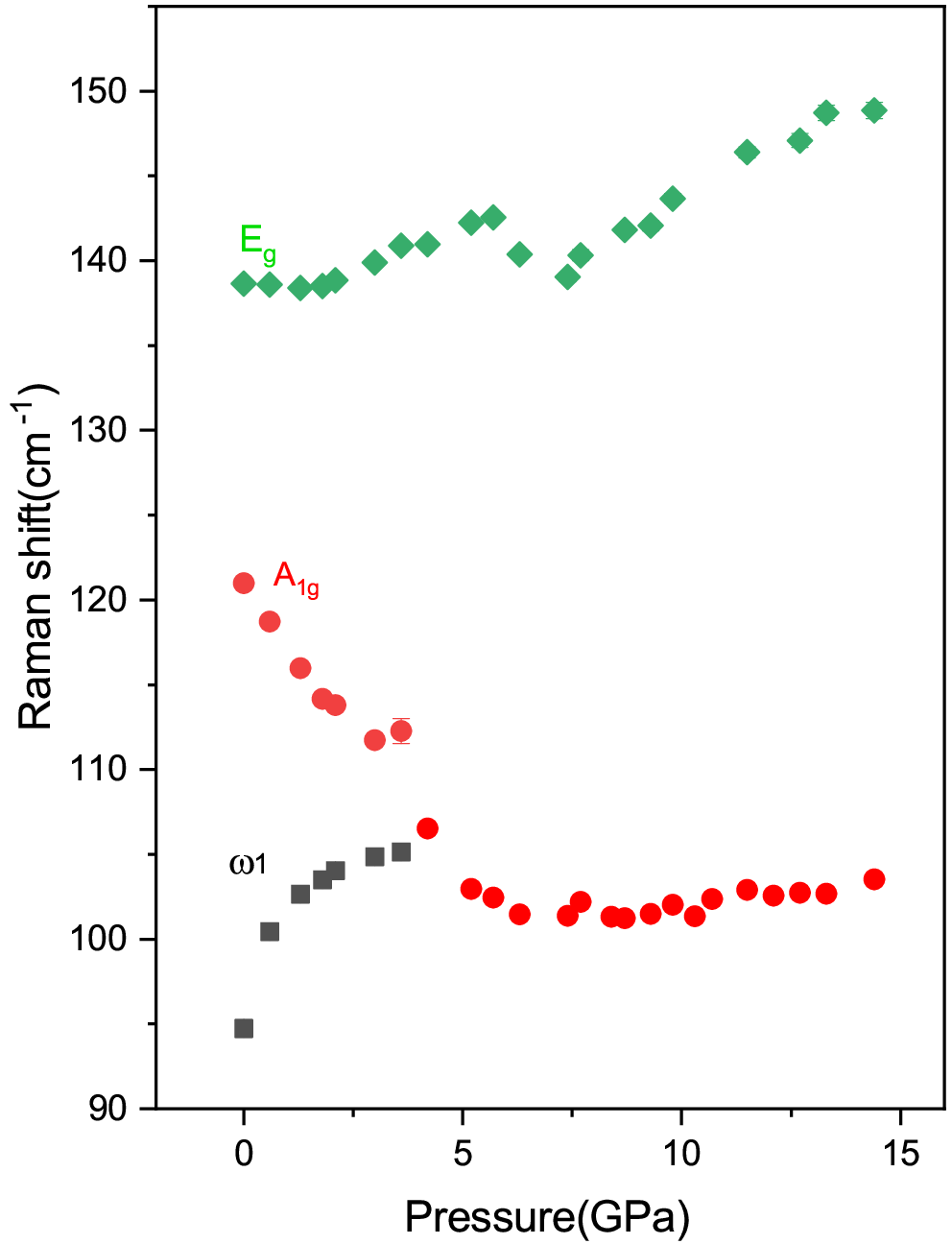}
		\caption{Variation of Raman Peak position with Pressure. The green symbol indicates the $E_g$ mode and the red and black symbols represents the $A_{1g}$ and $\omega_1$ mode respectively }
		\label{raman peak position}
	\end{figure}

\begin{figure}[ht!]
		\centering
		\includegraphics[width=0.65\linewidth]{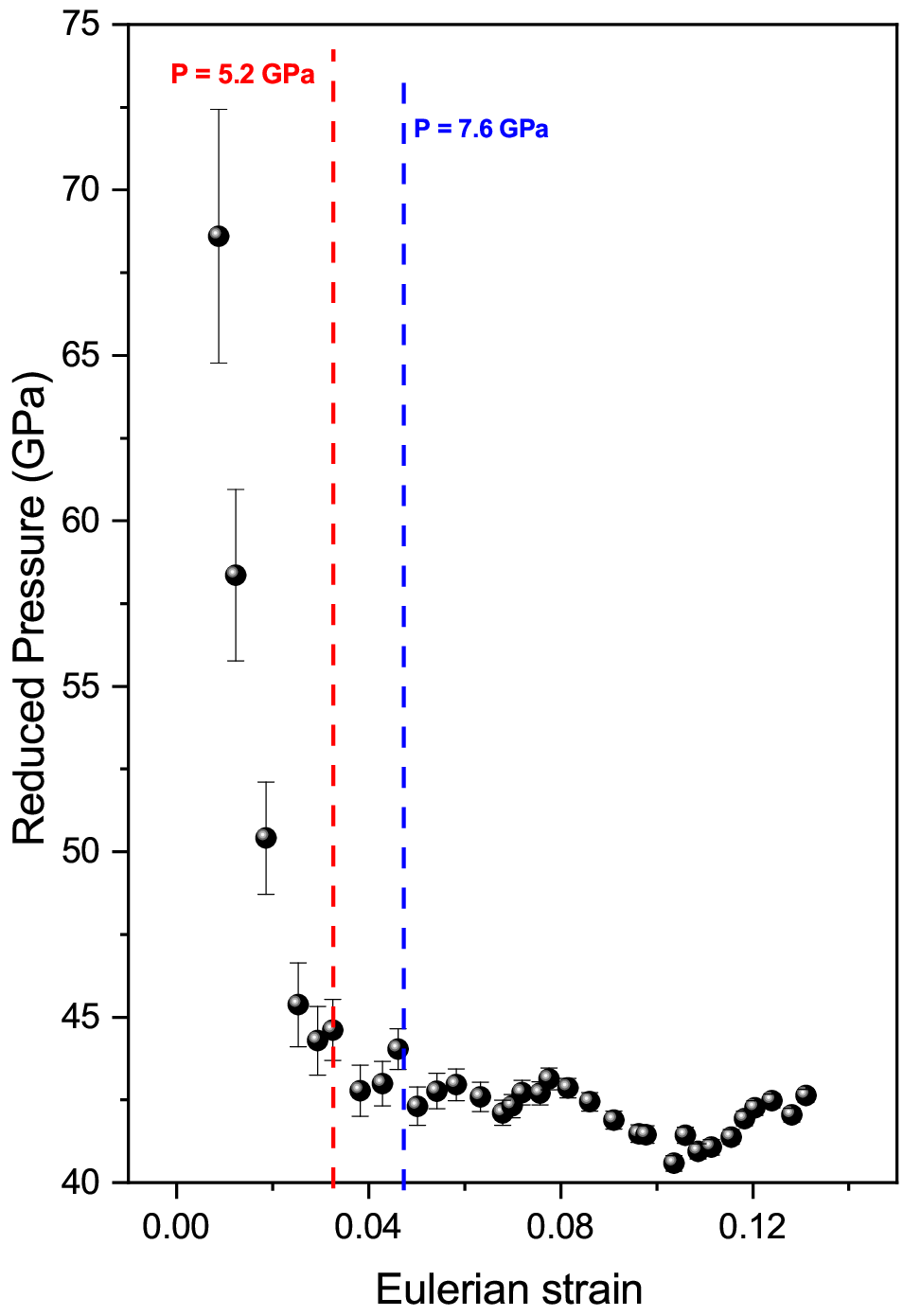}
		\caption{The reduced pressure vs Eulerian strain over the whole pressure region. The reduced pressure drops drastically up to 5.2 GPa, shown by the dashed red line. The dashed blue line corresponds to 7.6 GPa pressure, beyond which, the reduced pressure remains almost constant.  }
		\label{eulerian strain}
	\end{figure}

\begin{figure}[ht!]
		\centering
		\includegraphics[width=0.6\linewidth]{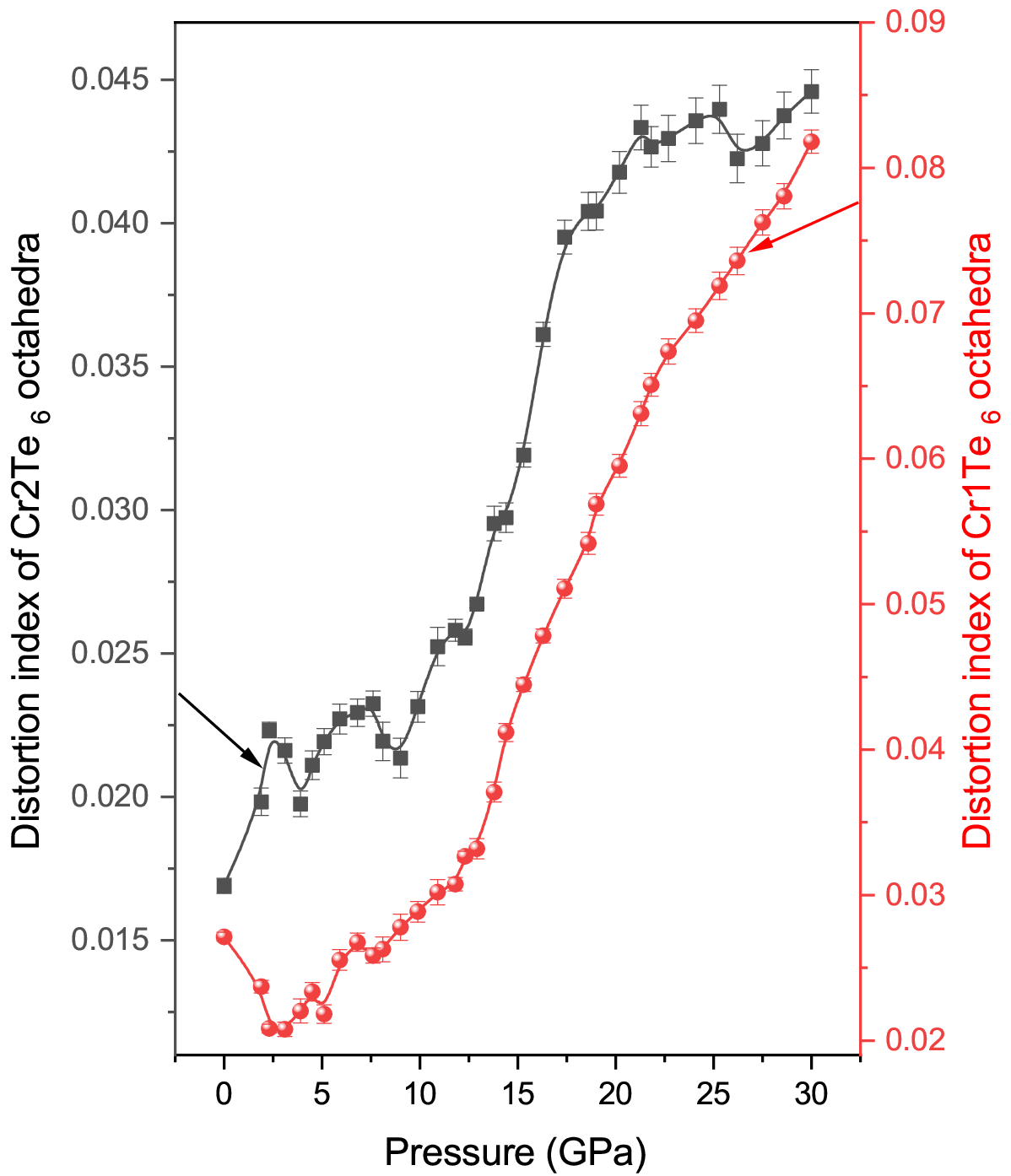} 
		\caption{The variation of distortion index (DI) with pressure. The red dots represent the DI of $Cr1Te_6$ octahedra and the black squares denote the DI of $Cr2Te_6$ octahedra. The red and black lines are a guide to the eye. The arrows denote the respective axis of the plots. }
		\label{distindex}
	\end{figure}

     We have plotted the Raman shift with pressure in Fig.~\ref{raman peak position}. It can be seen clearly that the $A_{1g}$ mode softens till about 7.6 GPa and then shows very slight hardening. $\omega_1$ mode increases initially and then softens slightly to merge with $A_{1g}$ mode. The Raman shift of $E_g$ mode shows a blue shift till about 5.7 GPa and then suddenly decreases till about 7.6 GPa. Above 7.6 GPa pressure, it again shows a blue shift with the increase in pressure. Softening of a Raman mode is generally associated with a structural transition. However, in XRD experiments, we do not encounter any structural change with pressure. Therefore, we have calculated the Eulerian strain behavior to see any changes in the internal strain of the lattice. In the absence of pressure-driven structural transition, the induced internal strain plays an important role in determining the electronic properties of materials \cite{PhysRevB.83.113106,saha2020pressure,jana2016high}. To investigate the impact of such strain on our sample, we have calculated the Eulerian strain and reduced pressure using the following relations\cite{PhysRevB.83.113106}
     \begin{equation}
     	H=\frac{P}{3f_E(1+2f_E)^{5/2}} 
     \end{equation}
     \begin{equation}
     	f_E=\frac{1}{2}*[(\frac{V_0}{V})^{\frac{2}{3}}-1]
     \end{equation}
     where H is the reduced pressure, $f_E$ is the Eulerian strain, and $V_0$ is the volume at ambient pressure. In the absence of any structural phase transition, the reduced pressure typically exhibits a linear relationship with Eulerian strain. However, in our case, we observe that the reduced pressure initially decreases drastically up to about 5.2 GPa and then becomes almost constant beyond 7.6 GPa as shown in Fig.~\ref{eulerian strain}. The decrease in Eulerian strain indicates an increase in ordering of the lattice with pressure. This can be seen in the drop in the values of the distortion index in both the octahedra, as shown in Fig.~\ref{distindex}. Therefore, the $\omega_1$ and the $A_{1g}$ mode, which were found to be split at ambient conditions, merge together above 5.2 GPa. In addition to the further softening of $A_{1g}$ mode till 7.6 GPa, $E_g$ mode shows a sudden drop at the same pressure. Now we move our attention to the disappearance of Raman modes above 15 GPa. The distortion index of both the octahedra increases rapidly above 10 GPa, which may lead to the large disorder in the formula units, and hence no sharp Raman modes are seen.

    In the absence of a structural phase transition, magnetic ordering may influence the Raman modes. This is supported by earlier observations of temperature-induced magnetic transitions in layered magnetic compound CrSBr, where a change in Raman mode was observed at the vicinity of paramagnetic to ferromagnetic phase transition and also at the vicinity of ferromagnetic to antiferromagnetic phase transition \cite{PhysRevB.107.075421}. In the ferromagnetic double perovskite $Lu_2NiMnO_6$ pressure-induced ferromagnetic to antiferromagnetic transition is observed with no structural changes \cite{terada2020pressure}. Therefore, in the present case, it is possible that pressure induces a change in the magnetic ground state, such as a transition from a ferromagnetic to a different magnetic configuration, which could account for the observed anomalies in the Raman spectra. But only from the Raman and XRD data we can not confirm the magnetic transition, so we have carried out  Density Functional Theory (DFT) analysis. But before going to the DFT part, it is important to see whether the Raman modes in our sample are magnetically active or not. In order to see the effect of ferromagnetic to antiferromagnetic transition on the Raman modes of $Cr_{3}Te_4$ we have performed low low-temperature Raman experiment from Room temperature to 22K.

 \subsection{Low temperature Raman}
\begin{figure}[ht!]
		\centering
		\includegraphics[width=0.6\linewidth]{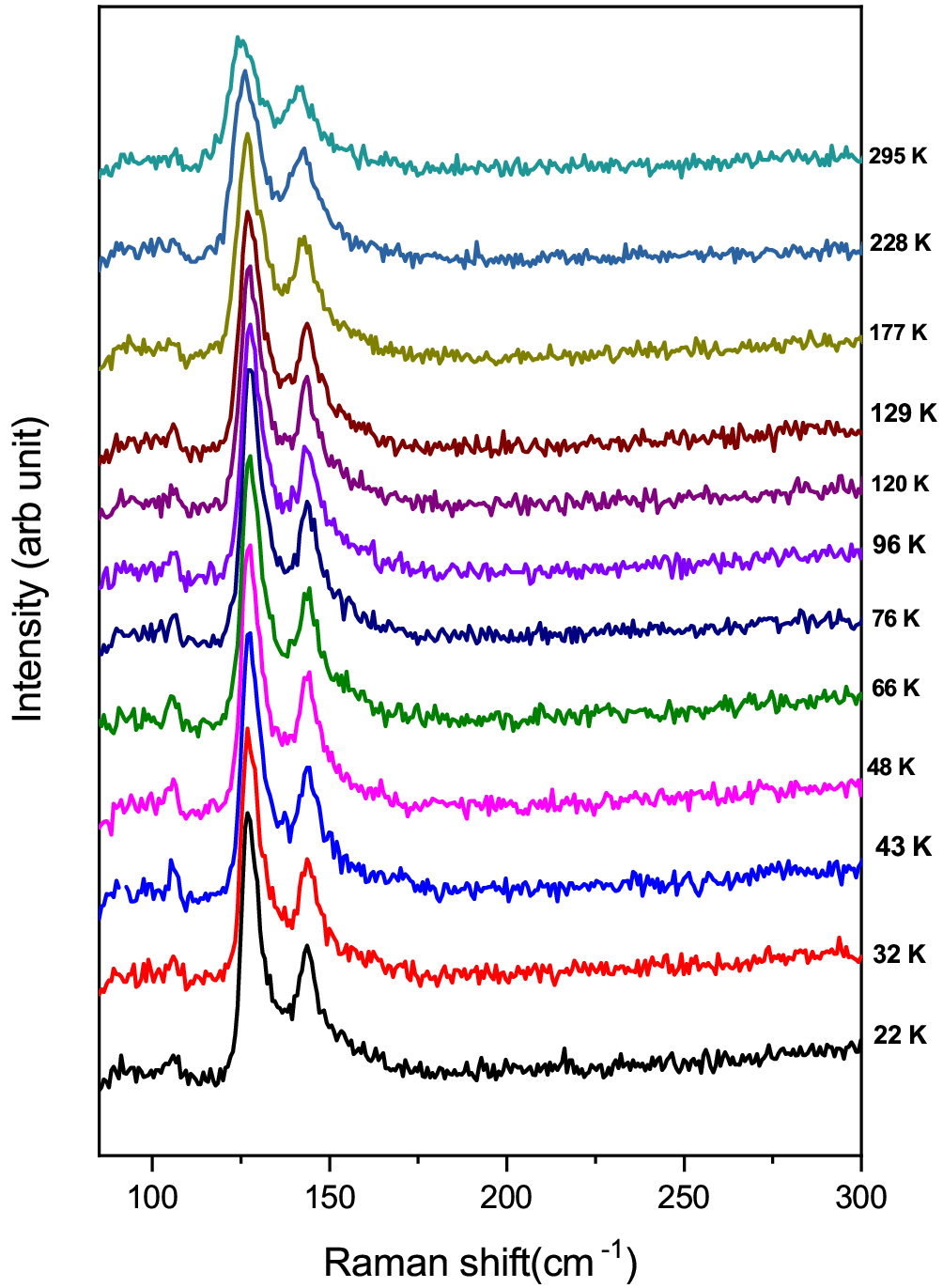} 
		\caption{The evolution of Raman spectra with temperature }
		\label{temp evolution of raman spectra}
	\end{figure}

 As mentioned earlier, $Cr_{3}Te_4$ is a room temperature ferromagnet, and below the Neel temperature $T_N \sim 90-110 K$ a canted antiferromagnetic state is observed. The evolution of Raman spectra with temperature at some selected temperature values is shown in Fig.~\ref{temp evolution of raman spectra}. We have not observed the emergence or disappearance of any new peaks in between the observed temperature range, suggesting the absence of any temperature-induced structural phase transition. All the Raman mode intensities were corrected using the Bose-Einstein(BE) thermal correction to account for the temperature-dependent vibrational state occupancy.
 \begin{equation}
 	I_{red} = I_{obs}* (1-exp(\frac{-hc\omega}{kT}))
 \end{equation}
 where $I_{red}$ is the Raman intensity after BE thermal correction, $I_{obs}$ is the observed intensity, $\omega$ is the Raman shift, T is the temperature, and K is the Boltzmann constant.
  First, we tried to fit both the peaks using a Lorentzian profile, but it did not give a good fit. There is a clear mismatch between the observed data points and the fitted curve, which can be seen in Fig.~S5(b) of the supplementary material. This mismatch can be due to coupling between the phonon and electronic continuum present in the system, giving a Fano-like asymmetry. So, we fitted the $A_{1g}$ mode using a Breit-Wigner-Fano (BWF) function and the $E_{g}$ mode using a Lorentzian profile which gives an excellent fit(Fig.~S4(a)). The BWF function is given by the following formula \cite{PhysRev.124.1866,PhysRevB.108.014438}
  \begin{equation}
  	I(\omega)= A\frac{(q+\epsilon)^{2}}{1+\epsilon^2} , \hspace{0.6 cm}  	\textit{where} ~\epsilon=\frac{\omega-\omega_c}{\Gamma}
  \end{equation} 
  $\omega$ is the Raman shift, $\omega_c$ is the Raman peak position, q is the coupling parameter, $\Gamma$ is the FWHM of the peak, and A is a constant. 
  
\begin{figure}[ht!]
	    \centering
	    \includegraphics[width=0.9\linewidth]{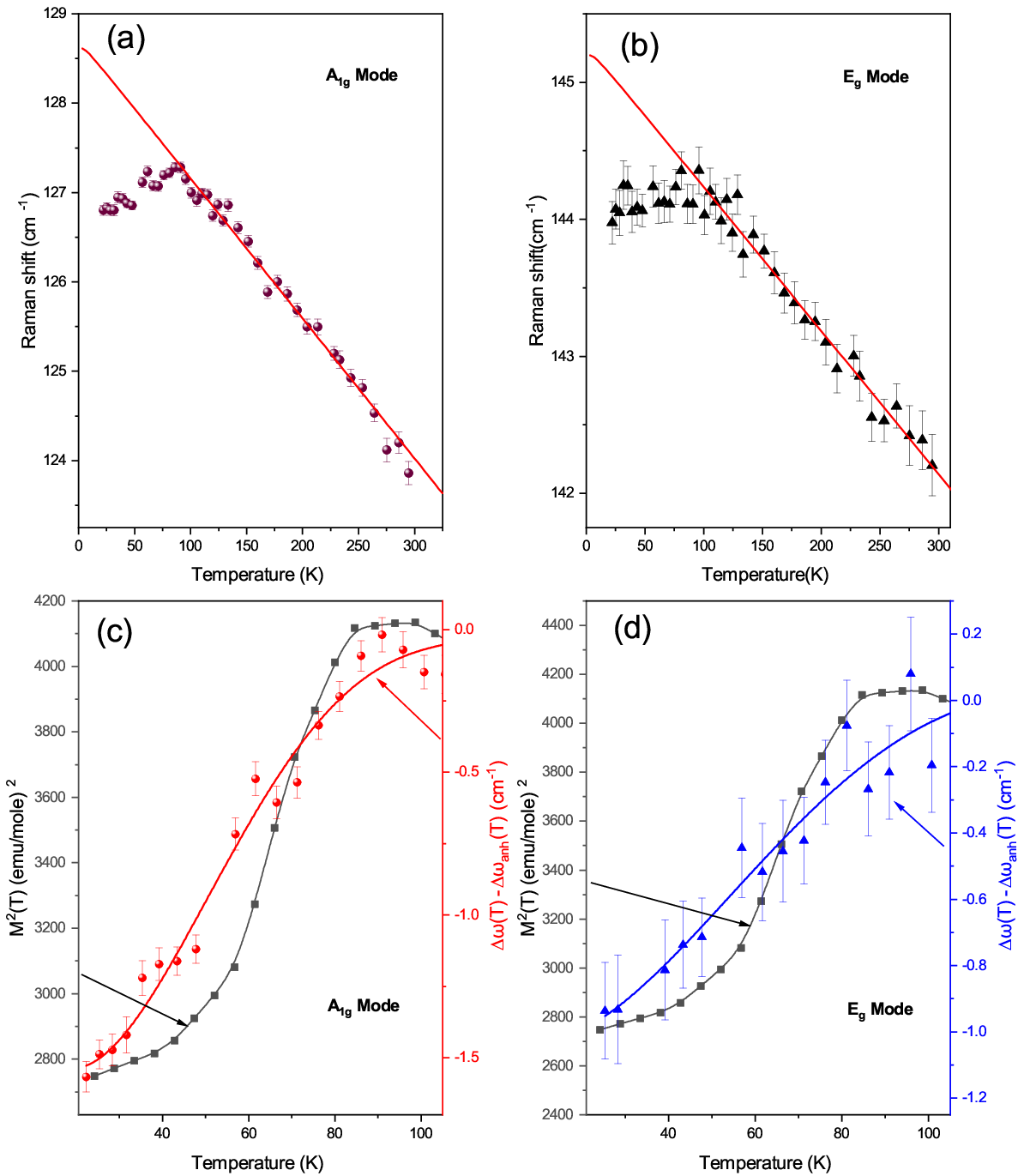}
		\caption{Temperature variation of Raman shift of (a) $A_{1g}$ mode, (b) $E_g$ mode. Red line in (a) and (b) represent the anharmonic model fit to Eq.\ref{anh} up to the Neel temperature. (c) and (d): show the high degree of correspondence between the $M^{2}(T)$ and the difference between observed $\omega(T)$ and fitted $\omega_{anh}(T)$ for $A_{1g}$ and $E_{g}$ mode respectively. The lines through the data points are guide to the eye.}
		\label{anharmonic_low temp}
		 
	\end{figure}

  As the temperature is lowered, both modes exhibit a conventional blue shift. However, a notable deviation from this trend is observed near 100 K, below which both modes exhibit anomalous softening. In the absence of any structural phase transition, this behavior is particularly interesting as it coincides with the magnetic ordering temperature of the system. In magnetic materials, the temperature dependence of the Raman mode frequency is governed by several factors, which are given below \cite{PhysRevB.60.11879}
  \begin{equation}
  	\Delta\omega(T)= \Delta\omega_{latt} + \Delta\omega_{anh} + \Delta\omega_{Sp-Ph} + \Delta\omega_{e-Ph}
  	\label{temp dependance of raman mode freq}
  \end{equation}
  With the increase in temperature, the crystal lattice undergoes expansion, which contributes to the shift in Raman mode frequencies. As volume expands, the force constant that governs the atomic vibration decreases, and hence Raman shift $\omega$ decreases.  In our sample, the change in unit cell volume with temperature is only $0.3~ \%$ \cite{doi:10.1021/acs.inorgchem.2c01826}, so we can safely ignore the volume dependence term in Eq.\ref{temp dependence of raman mode freq}. When the temperature increases, the anharmonic lattice vibration also contributes to the shift in Raman frequency. In a three-phonon decay process, one optical phonon decays into two acoustic phonons of same frequency \cite{PhysRevB.28.1928}. This is given by Eq.\eqref{anh}
  \begin{equation}
  	\Delta\omega_{anh}(T)= \omega_0 - B(1+\frac{2}{e^{\phi/2}-1}), \hspace{0.6 cm} \phi = \frac{\hbar\omega_0}{K_BT}
  	\label{anh}
  \end{equation}
  First, we tried to fit the whole temperature range using Eq.\eqref{anh}, but it did not give a good fit, so we fitted it up to 100 K and then extrapolated it as shown in Fig.~\ref{anharmonic_low temp}. A noticeable deviation between the fitted anharmonic model and the experimental data is observed below 100 K for both Raman modes. This discrepancy indicates the presence of spin-phonon coupling interaction, which is not captured by the pure anharmonic model. The interaction between the electron spin and phonon affects the Raman mode frequency as reported by Baltensperger and Helman \cite{baltensperger1970influence}. This effect can be best understood by the following simplified formula\cite{PhysRevB.108.014438}
  \begin{equation}
  	\Delta\omega(T) - \Delta\omega_{anh}(T) = F* |M(T)|^2
  	\label{sp-ph}
  \end{equation}
  F is a constant, and M(T) is the magnetization as a function of temperature. The M(T) data is obtained as a private communication with Thirupathaiah et. al and will be published somewhere else.  We have plotted the Eq.\eqref{sp-ph} in the Fig.~\ref{anharmonic_low temp}(c) and (d) and found a high degree of correspondence, which indicates the presence of strong spin-phonon coupling in the system. 
  \subsection{Computational study} 
\begin{figure}[ht!]
		\centering
		\includegraphics[width=1.0\linewidth]{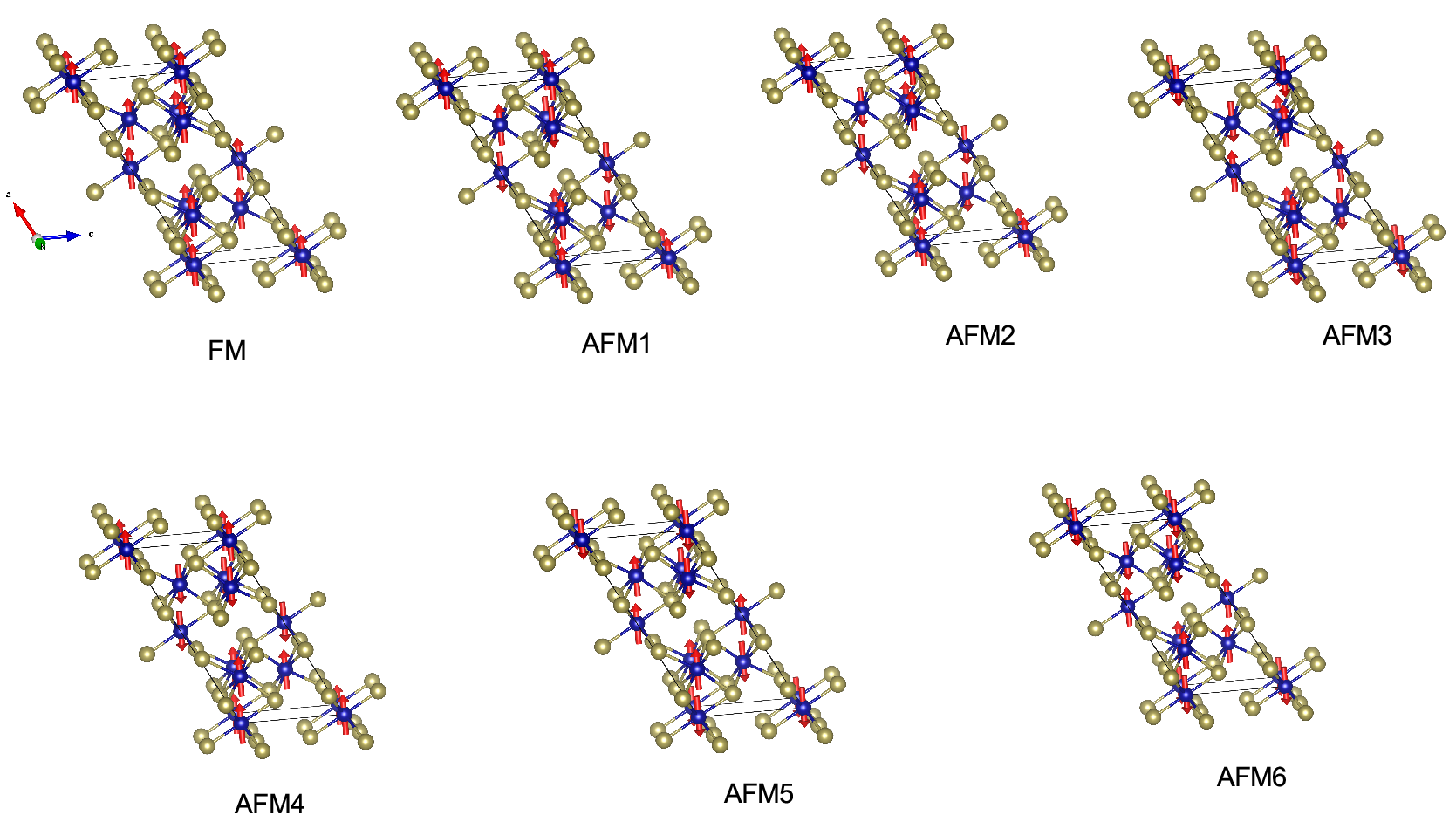}

		\caption{ All the considered magnetic configuration. The FM stands for ferromagnetic configuration and the AFM stands for antiferromagnetic configuration. AFM2 is the most stable antiferromagnetic configuration. The blue circle in the figure represents the Cr atom and the golden circle represents the Te atom. }
		\label{mag_configuration}
	\end{figure}

\begin{figure}[ht!]
		\centering
		\includegraphics[width=0.6\linewidth]{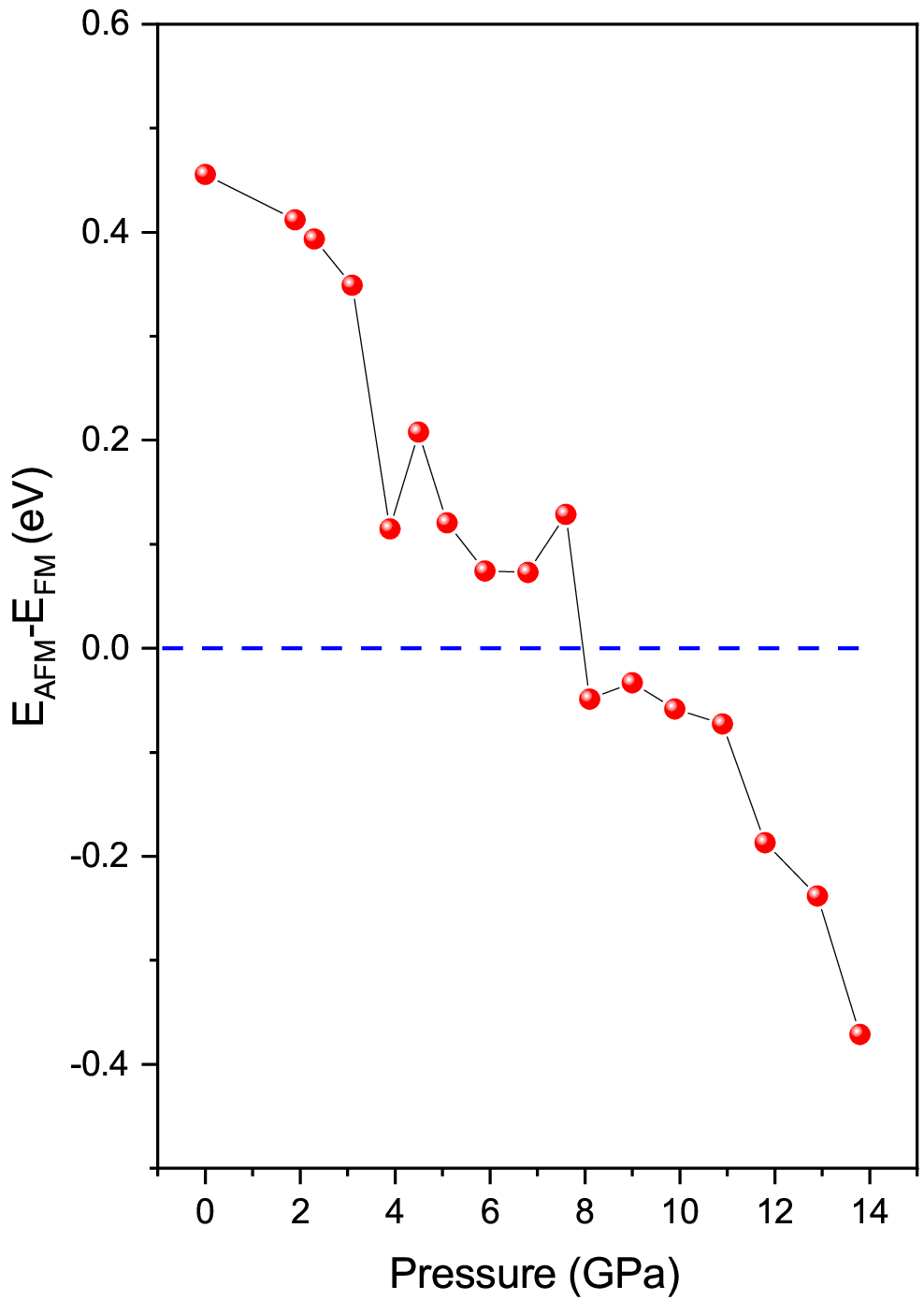} 
		\caption{The Variation of $E_{AFM}-E_{FM}$ with pressure. The red circle represents the calculated data and the blue dotted line corresponds to the 0 value of $E_{AFM}-E_{FM}$ indicating the FM to AFM transition }
		\label{AFM-FM}
	\end{figure}

   To understand the effect of external pressure on the magnetic ground state of the sample, we have carried out detailed DFT  calculations. Based on symmetry, we have considered 6 antiferromagnetic(AFM), 1 ferromagnetic(FM), and 1 non-magnetic (NM) configuration. Out of all the possible configurations, the ferromagnetic ground state is the most stable with total magnetic moment $\mu= 19.9943~{\mu_B}$. The magnetic moment of $Cr_1$ atom is $3.336~\mu_B$ and $Cr_2$ atom is $3.324~\mu_B$, which matches well with the previous study \cite{dijkstra1989band}.  Based on the atomic arrangement of Cr, we identified 6 AFM configurations. We also considered other possible AFM configurations, which are not shown here. However, they did not converge to perfect AFM states (i.e $\mu$ $\neq$ 0) and were found to be energetically unstable.   All the six possible AFM configurations are shown in Fig.~\ref{mag_configuration}. The AFM2 represents the most stable AFM configuration. At each pressure point, we compared the ground state energy of FM and AFM configurations and found that the ferromagnetic ground state remains energetically stable up to 7.6 GPa. Above 7.6 GPa the AFM ground state becomes stable.  The energy difference between the AFM and FM states, plotted as a function of pressure (Fig.~\ref{AFM-FM}), remains positive up to 7.6 GPa, indicating FM stability. Above this pressure, the energy difference becomes negative, signifying a transition to an AFM ground state. The energies of all the considered magnetic configurations at ambient pressure and at 9 GPa pressure are given on table~\ref{table_2}. The energy of the stable configuration is taken as reference.

   \begin{table}[ht!]
   	\addtolength{\tabcolsep}{-9.5pt}
   	
   	\begin{center}
   		\begin{tabular}{ |c|c|c|c|c|c|c|c|c|} 
   			\hline
   			 & \multicolumn{8}{c|}{Energy(eV/formula unit)}  \\ \cline{2-9}
   			 
   				Pressure(GPa)&FM&AFM1&AFM2&AFM3&AFM4&AFM5&AFM6&NM\\
   			\hline
   			0&0&0.2096335&0.227809&0.183704&0.1422155&0.2096335&0.1422155&3.2151735\\
   			\hline
   			9&0.016623&0.0239105&0&0.069413&0.0141825&0.0239105&0.0141825&1.533542\\
   			\hline

   		\end{tabular}\\
   	\end{center}
   	\caption{The energy (eV/formula unit) of all the considered magnetic configurations at 0 GPa and 9 GPa pressure. At 0 GPa pressure, the FM configuration is stable and the energy of other configurations are written by considering the energy of the FM configuration as reference. At 9 GPa pressure, the AFM2 configuration is stable and considered as the reference.}
   	\label{table_2}
   \end{table}

\begin{figure}[ht!]
		\centering
		\includegraphics[width=1.0\linewidth]{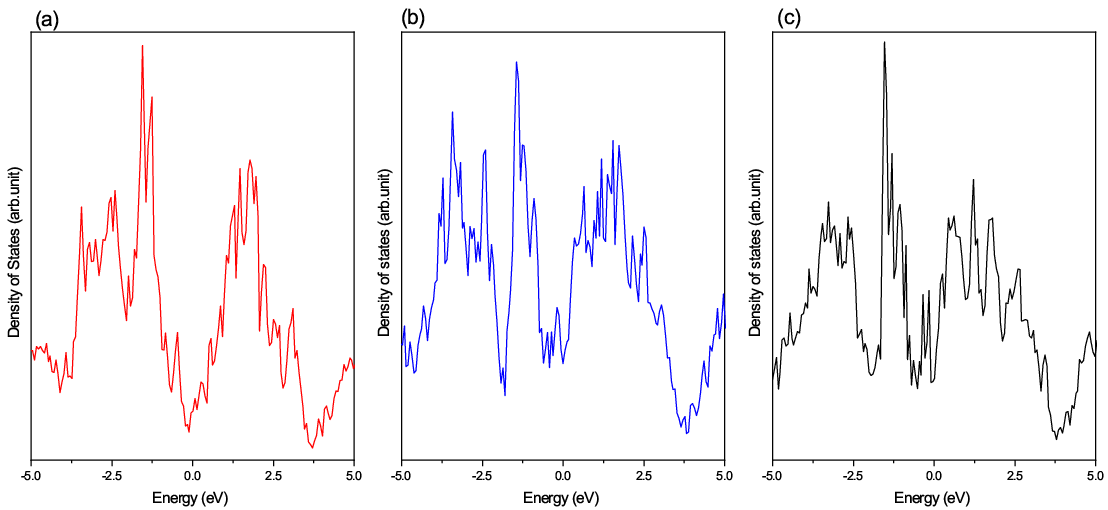}
		\caption{(a) The density of state of ferromagnetic $Cr_{3}Te_4$ at 0 GPa pressure (b) The density of state of antiferromagnetic $Cr_{3}Te_4$ at 9 GPa pressure (c) The density of state of antiferromagnetic $Cr_{3}Te_4$ at 13.8 GPa pressure     }
		\label{dos}
		
	\end{figure}

   Now, in order to see the effect of this magnetic transition on the band structure of $Cr_{3}Te_4$, we have calculated the density of state (DOS) for the ambient ferromagnetic phase and also for the antiferromagnetic phase at 9 GPa and 13.8 GPa as shown in Fig.~\ref{dos}. The continuous DOS at the Fermi level for all the pressure values reveals the metallic nature of the sample throughout the pressure values, which excludes the possibility of an electronic transition. 
    
   This pressure-induced ferromagnetic to antiferromagnetic transition is corroborated by our Raman spectroscopy data, where we observe a sudden drop in $E_g$ mode around 7.6 GPa and  the softening of $A_{1g}$ mode till about 7.6 GPa. So the observed anomalies in the Raman data around 7.6 GPa can be attributed to a pressure-induced FM to AFM transition. 
   
   The magnetic properties in $Cr_{3}Te_4$ arise due to a combination of direct exchange, super exchange, and double exchange interaction. At ambient condition the super exchange interaction between the intralayerd $Cr_{1}-Te-Cr_{1}$ and $Cr_{2}-Te-Cr_{2}$, as well as the double exchange interaction between the interlayered $Cr_{1}-Te-Cr_{2}$ are responsible for the stabilization of a ferromagnetic ground state. The direct exchange interaction between the $Cr_{1}-Cr_{1}$ and $Cr_{2}-Cr_{2}$ favours the antiferromagnetic state \cite{wang2023field}. Beyond 8 GPa Pressure, the c axis becomes more compressible as compared to the a and b axes as shown in Fig.~S1 of the supplementary material. The unit cell of $Cr_{3}Te_4$ comprises of alternating Cr filled and Cr vacant layers along the c direction. The increased compressibilty along the c direction reduces the intralayer separation and thereby brings the magnetic Cr atoms closer to each other. This anisotropic compression along the c direction enhances the direct exchange interaction, making it dominant over the competing superexchange and double exchange interaction and leading to a pressure-induced transition to the antiferromagnetic ground state. The observed anomalies in the Raman spectroscopy data near the magnetic ordering pressure reveal the strong correlation between the spin ordering and lattice vibration.
   \subsection{Conclusions}  
   In conclusion, the detailed high-pressure behavior of $Cr_{3}Te_4$ has been thoroughly investigated using X-ray diffraction, Raman spectroscopy, and the first-principle DFT calculations. The monoclinic structure remains stable up to the highest pressure. The anomalies observed in the Cr-Te bond length and octahedral volume indicate to a pressure-induced magnetic phase transition. This is supported by the complementary Raman spectroscopic measurements, where the $A_{1g}$ mode softens till about 7.6 GPa and then hardens very slightly. The $E_g$ mode also shows a sharp drop around the same pressure region. The low temperature Raman data reveals the softening of both $A_{1g}$ and $E_{g}$ Raman modes at the onset of ferromagnetic to antiferromagnetic transition temperature. The presence of strong spin-phonon coupling is also revealed. The DFT calculations confirm that the observed anomalies in the Raman modes around 7.6 GPa are related to the pressure-induced switching of ferromagnetic to antiferromagnetic ground state. The anisotropic compression along the c direction around that pressure makes the direct exchange interaction dominant, resulting in a pressure-induced antiferromagnetic ground state. These findings reveal the pressure-induced modification in the magnetic structure and the presence of magneto-structural coupling in the sample, which opens the pathway for future applications.
   \section{Acknowledgments}  
   Portions of this research were carried out at the light source PETRA-III of DESY, a member of the Helmholtz Association(HGF). Financial support by the Department of Science and Technology (Government of India) provided within the framework of the India@DESY collaboration is gratefully acknowledged. AKM acknowledges the IISER Kolkata, for the
   financial support and experimental facilities to carry out the PhD work. S.T. thanks Anusandhan National Research Foundation (ANRF), India, for the financial support through Grant no. CRG/2023/00748. This research has used the Technical Research Centre (TRC) Instrument facilities of S. N. Bose National Centre for Basic Sciences, established under the TRC project of the Department of Science and Technology (DST), Govt. of India.

 	\bibliography{mybib}
	\bibliographystyle{unsrt}
	
	\newpage
	\section{Figures}
	
	

	

\end{document}